# Using Layout Information for Spreadsheet Visualization

Sabine Hipfl
Institut für Informatik-Systeme
Universität Klagenfurt
shipfl@edu.uni-klu.ac.at

**ABSTRACT**

*This paper extends the spreadsheet visualization technique reported in [Clermont, 2003] by using layout information. The original approach [MC, 2002] identifies logically or semantically related cells by relying exclusively on the content of cells for identifying semantic classes.*

*A disadvantage of semantic classes is that users have to supply parameters which describe the possible shapes of these blocks. The correct parametrization requires a certain degree of experience and is thus not suitable for untrained users. To avoid this contstraint, the approach reported in this paper uses row-/column-labels as well as common format information for locating areas with common, recurring semantics. Heuristics are provided to distinguish between cell groups with intended common semantics and cell groups related in an ad-hoc manner.*

## 1 INTRODUCTION

Systematic research and field audits [Panko, 1998] have shown that erroneous spreadsheet programs are wide spread, even as base for important decisions. This is due to lack of systematic development and testing. To overcome the poor quality of spreadsheets, in recent years, some testing (see [RLDB, 1998], [RRB, 1999] or [AC, 2002]) and visualization methods (see [Clermont, 2003], [Sajaniemi, 2000] or [Butler, 2000]) for spreadsheets have been developed.

Model visualization, as developed by Clermont [Clermont, 2003], consists of three different methods to aggregate semantically related cells to groups in order to get an abstraction of the spreadsheet and to build a comprehensive representation as basis for detecting errors. One of these methods serves to detect repeating components of spreadsheets, called semantic classes [MC, 2002]. The approach reported in this paper extends this approach by including layout information.

Semantic classes are mainly based on a notion of similarity between cells, so called *logical equivalence classes* [CHM, 2002]. These equivalence classes were initially used as a base for grouping cells with the same content into *logical areas* [ACM, 2000]. This approach was extended to assign similar groups of cells, so called *semantic units*, into *semantic classes* [MC, 2002]. By this, semantic classes are an abstraction of a spreadsheet, based on the similarity of formulas. The user has to specify, whether semantically related cells are spread out column-wise, row-wise or block-wise. Thus the content of cells and spatial situations are taken into account and it is checked, whether the semantic content of these areas is of repetitive nature.

The spatial parametrization can be tricky in certain cases, as the orientation can be indicated only for the whole spreadsheet. Sometimes a spreadsheet consists of different blocks of cells more or less related. As the orientation of one block could be row-wise, that of the other block column-wise. The parametrization must be done by the spreadsheet user itself. As this demands a high knowledge of the structure of the spreadsheet it requires a certain degree of experience and might not be suitable for the untrained user.

Moreover, semantic classes only identify semantically related blocks of cells having the same extent. This requires the blocks to be of the same size. But blocks of cells that are semantically related but do have a different size will not be identified in their full extent.

The approach presented in this paper overcomes these hurdles, as explicit parametrization is avoided. This is done by using layout information to identify and group semantically related cells, yielding so called *layout areas*. These can then be checked whether they contain semantic classes.

Layout information in spreadsheets is, for example, bands of labels or frames. The spreadsheet-programmer uses layout information to structure the spreadsheet and to increase the understandability and readability of the sheet. The main idea behind using layout information for spreadsheet visualization is based on the general area of application of layout information: the spreadsheet-programmer uses layout information to describe semantically related areas on the spreadsheet. For the reconstruction of these related areas layout information is used again. By this

- explicit parametrization is avoided as
- implicit identification of the orientation of cells is possible,
- the orientation is identified for each block of cells individually and
- it is possible to identify semantically related blocks of cells with different extent.

Including formulas, spatial situations and layout information of cells can help

- to get a better view of the spreadsheet and
- to better recognize related areas.

This can lead to two major advantages:

- The comprehension of the whole spreadsheet improves and
- the reconstructed model of the spreadsheet is more meaningful.

Briefly, the technique that is subsequently presented assigns cells to labels and then detects similarities between cells that are assigned to similar labels. The notion of similarity between groups of cells that is used throughout this paper is based on logical equivalence [MC, 2002]. The generated abstraction consists of semantic classes where each semantic unit bears a label. Thus, the auditing techniques that can be applied to this abstraction are the same as [CM, 2003] suggest for semantic classes.

In contrast to existing techniques spreadsheet-programmers are not forced to introduce any specific layout characteristics into a spreadsheet. However, the approach is able to use the present information to create a meaningful abstraction of the spreadsheet. The analysis for similarity between the groups of cells goes also much more into detail as many of the commercial available toolkits do. Therefore, a high number of semantically equivalent blocks can result and color coding of similar cells of the spreadsheet is not directly supported. This is not a disadvantage at all because users are enabled to audit the spreadsheet on the base of the high level abstraction and not on the cell level, making the approach suitable for large spreadsheets.

The following section introduces different kinds of layout information and their relevance for the method of layout areas. In the third section some basic formal definitions are given, which are important for the implementation of the toolkit. In the fourth section the four heuristics on which layout areas are built are presented and illustrated by examples.

## 2 LAYOUT INFORMATION IN SPREADSHEETS

Layout information can be classified into three main groups:

- Labels
- Formatting of the content of cells
- Lines and Frames

In general, layout information is used to structure the spreadsheet. This makes the spreadsheet more readable as well as more understandable.

### 2.1 Labels

Labels are used to describe the semantics of cells and to help the spreadsheet-user to understand the usage of the whole spreadsheet, of blocks of cells as well as of single cells. Labels have the highest impact for the here introduced method of layout areas.

**Extent of Labels**

In general, labels can run horizontally to the right or vertically downwards. If labels are used to describe columns, they are called horizontal labels as they are running horizontally to the right. If labels are used to describe rows, they are called vertical labels as they are running vertically downwards.

The extent of the labels within a spreadsheet defines an implicit identification of the orientation of the spreadsheet or of a particularly labelled portion of the sheet. As horizontal labels are describing columns, such parts of a spreadsheet are column-oriented and as vertical labels are describing rows, such parts of a spreadsheet are row-oriented.

**Kinds of Labels**

The main idea of using labels for layout areas is the assumption that similar labels are an indicator for semantically related cells. As there are different degrees of similarity between labels, five kinds of labels can be defined. A formal definition of each kind of label is given in the next section.

- *Running numbers*
  In general, labels are cells containing a string. Running numbers are a special form of labels as they contain numbers. They are a set of cells which are running horizontally to the right or vertically downwards, containing values which are increasing in relation to the preceding cell at a given step size, usually one.

  Using running numbers in the same way as conventional labels makes sense, because in many spreadsheets running numbers are not used as numbers but as labels to describe blocks of cells. For example, the years *2001* to *2004* can be used to describe blocks of cells in which the cash flow is calculated for four years.

- *Labels interpreted as ordinal numbers*
  These kinds of labels are a special form of running numbers. They are not similar to each other. Hence a function or a dictionary is needed to map these labels to numbers or help from the user to identify this mapping.

  For example, the twelve months *January, February, March … December* can be mapped to the numbers *1 … 12*.

- *Counters*
  These kinds of labels are made up of one word with an attached running number in front or back of the word. It doesn't matter if there is a blank between the word and the running number.

  For example, the labels *Year 1, Year 2, Year 3* and so on are counters.

- *Labels with complete identity*
  Labels containing exactly the same word/s are called labels with complete identity.

- *Labels with partial identity*
  A set of labels that are made up of $n$ words ($n \geq 2$) are called labels with partial identity, if the first $m$ ($m \leq n-1$) words or the last $m$ ($m \leq n-1$) words are completely identical.

  For example, the labels *Cash flow to Present Value* and *Cash flow increment* are partially identical as they contain the completely identical words *Cash flow*.

**2.2 Formatting of the Content of Cells**

In spreadsheets different formatting of the content of cells is used to highlight related cells or to separate important issues from less important ones. The basic idea behind using different formatting for the identification of layout areas is the assumption that cells with the same format are assumed to be semantically related. Hence, format information is also used for layout areas.

Usually, there are five different kinds of formatting elements applicable for rendering the content of a cell: the *font* itself, the *font size*, the *style of the font*, the *font color* and finally the *background color* of a cell. In this work only the first four elements are used, the *background color* is left out.

**2.3 Lines and Frames**

In spreadsheets, lines and frames are used to separate cells, or more common, blocks of cells. This kind of layout information is not used for the identification of layout areas because the information it provides is ambiguous due to the problem of proper attribution. E.g., a horizontal line can belong to either one of the two cells that it separates. Because of this ambiguity, lines and frames are not considered. It might be worth consideration for further work.

**3 FORMAL DEFINITIONS**

As a base for implementation of a toolkit, some formal definitions have to be introduced. These definitions are based on the formal definitions in [Clermont, 2003].

In the following, $C$ denotes the set of all non empty cells of a spreadsheet, with $c \in C$: $((x, y), v, f, fi, ri)$. $(x, y)$ is the address of a cell, $v$ is the value, $f$ is the formula of cell (which can be empty), $fi$ is the formatting-information of a cell and $ri$ is the frame-information of a cell. The term *label* denotes a cell that contains a string and does not have a formula.

The definitions are stated in *Z*-syntax [Diller, 1996].

**Definition 1: Running Numbers**

Two cells $c_1, c_2 \in C$ are running numbers, if

- $c_1 = ((x_1, y_1), v_1, f_1, fi_1, ri_1) \land c_2 = ((x_2, y_2), v_1 + 1, f_2, fi_2, ri_2) \land$
- $( (x_1 = x_2 \land y_1 < y_2 \land \nexists c_3 = ((x_3, y_3), v_3, f_3, fi_3, ri_3)) \in C \mid x_3 = x_1 \land y_1 < y_3 < y_2) \lor$
- $(y_1 = y_2 \land x_1 < x_2 \land \nexists c_3 = ((x_3, y_3), v_3, f_3, fi_3, ri_3)) \in C \mid y_3 = y_1 \land x_1 < x_3 < x_2) )$

These properties guarantee that

- the value in cell $c_2$ is the value in $c_1$ plus one,
- if the running numbers are arranged vertically no other cell is between them or
- if the running numbers are arranged horizontally no other cell is between them.

**Definition 2: Labels interpreted as ordinal numbers**

A label $l \in C$, with $l = ((x, y), v, f, fi, ri)$, is interpretable as ordinal number, if there is an injective function $f$: Strings' $\rightarrow \{0, 1, \ldots, n\}$, with Strings' $\subseteq$ Strings and $v \in$ Strings'.

**Definition 3: Counters**

Two labels $l_1, l_2 \in C$ are counters, if they fulfill the following requirements:

- $l_1 = ((x_1, y_1), v_1, f_1, fi_1, ri_1) \land l_2 = ((x_2, y_2), v_2, f_2, fi_2, ri_2) \land$
- $\exists r \in$ Strings $\mid (v_1 = r \circ n_1 \land v_2 = r \circ n_2) \lor (v_1 = n_1 \circ r \land v_2 = n_2 \circ r) \land$
- $c_1 = ((x_1, y_1), n_1, f_1, fi_1, ri_1) \land c_2 = ((x_2, y_2), n_2, f_2, fi_2, ri_2)$ are running numbers.

These properties guarantee that

- the values $v_1$ and $v_2$ are strings consisting of two parts:
- a word which is exactly the same in both labels and
- a running number in front or in back of the word.

**Definition 4: Complete Identity**

Two labels $l_1, l_2$ with $l_1 = ((x_1, y_1), v_1, f_1, fi_1, ri_1)$ and $l_2 = ((x_2, y_2), v_2, f_2, fi_2, ri_2)$ are completely identical if $v_1 = v_2$.

**Definition 5: Partial Identity**

Two labels $l_1$, $l_2$ with $l_1 = ((x_1, y_1), v_1, f_1, fi_1, ri_1)$ and $l_2 = ((x_2, y_2), v_2, f_2, fi_2, ri_2)$ are partial identical if
$\exists u, v, w \in \text{Strings} \mid (v_1 = u \circ w \land v_2 = v \circ w) \lor (v_1 = w \circ u \land v_2 = w \circ v)$.

## 4 LAYOUT AREAS

The basic idea behind layout areas is to assign cells to labels. There is a distinction between geometrical and semantical assignments. Definitions 6 and 7 introduce these two concepts.

### 4.1 Definitions

In the sequel, $L$ denotes the set of all labels of a spreadsheet. The term *label* denotes one of the five kinds of labels mentioned above and the term *cell* denotes a cell containing a formula or a numeric value that is used in computations.

Geometrical assignment is used to assign cells to labels based on specific geometrical restrictions.

### Definition 6: Geometrical Assignment (GA)

The set of geometrical assignable cells, with $c_i \in C$, of a label $l_1 \in L$ is defined as:

$GA_{l1} = \{c \in C \mid c = ((x_z, y_z), v_z, f_z, fi_z, ri_z) \land$
$\qquad l_1 = ((x_1, y_1), v_1, f_1, fi_1, ri_1) \land$
$\qquad ( (x_1 = x_z \land y_1 < y_z \land \nexists\, l_2 \in L \mid l_2 = ((x_2, y_2), v_2, f_2, fi_2, ri_2) \land$
$\qquad\qquad\qquad\qquad\qquad x_1 = x_2 \land y_1 < y_2 < y_z)$
$\qquad \lor$
$\qquad (y_1 = y_z \land x_1 < x_z \land \nexists\, l_2 \in L \mid l_2 = ((x_2, y_2), v_2, f_2, fi_2, ri_2) \land$
$\qquad\qquad\qquad\qquad\qquad y_1 = y_2 \land x_1 < x_2 < x_z) )$
$\qquad \land$
$\qquad v_z \notin \text{Strings}\}$

if:
$\forall\, c_1, c_2 \in GA_{l1} \mid dense(GA_{l1}, c_1, c_2, d)$

These properties guarantee that a cell is geometrically assigned to a label only if

- the label is above or on the left side of the cell,
- there is no other label between the cell and the label and
- the cell itself is not a label.

The predicate *dense* is true, if either $c_1$ is separated only by $d$ cells from $c_2$, or there exists another cell $c_3 \in GA_{l1}$, such that $dense(GA_{l1}, c_1, c_3, d)$ and $dense(GA_{l1}, c_3, c_2, d)$. A more formal definition of *dense* is given in [MC, 2002].

It is important to mention that a cell can only be assigned to one label. If there are two labels that a cell could be assigned to, a horizontal and a vertical one, two different cases are possible:

- If the two labels are of a different kind the problem is solved by heuristics 1 and 2.
- If both labels are of the same kind the labels are equivalent and it doesn't matter to which label the cells are assigned.

In contrast to other approaches the user has only to indicate the maximum allowed number of empty cells within a block of cells.

The semantical assignment is used to assign cells to labels based on the formatting information of the content of the cells and labels.

**Definition 7: Semantical Assignment (SA)**
The set of semantically assignable cells, with $c_i \in C$, of a label $l \in L$ is defined as:

$$\begin{aligned}
SA_l = \{c \in C \mid & c = ((x_z, y_z), v_z, f_z, fi_z, ri_z) \wedge \\
& l = ((x_1, y_1), v_1, f_1, fi_1, ri_1) \wedge \\
& (fi_l = fi_z \vee v_z = \textit{Undefined}) \wedge \\
& ( \, ( \nexists \; c_2 \in C \mid c_2 = ((x_2, y_2), v_2, f_2, fi_2, ri_2) \wedge \\
& \qquad\qquad x_1 = x_2 \wedge y_1 < y_2 < y_z \wedge \\
& \qquad\qquad fi_z \neq fi_2) \\
& \quad \vee \\
& \quad ( \nexists \; c_2 \in C \mid c_2 = ((x_2, y_2), v_2, f_2, fi_2, ri_2) \wedge \\
& \qquad\qquad y_1 = y_2 \wedge x_1 < x_2 < x_z \wedge \\
& \qquad\qquad fi_z \neq fi_2) \, ) \\
& \wedge \\
& c \in GA_l \}
\end{aligned}$$

These properties guarantee that the semantical assignment of a cell to a label is only done if

- the label and the content of the cell are having the same formatting information or the value of the cell itself is not defined (which means empty),
- there is no other cell between the cell and the label, containing another kind of formatting information and
- the cell must be geometrically assignable to the label, as expressed by $c \in GA_l$.

As the same in geometrical assignment, the assignment is only done, if the cell is not assignable to another label.

**Definition 8: Layout Area**
A layout area is set of cells $L$ and a label $l$, if $c \in L \Rightarrow (c \in GA_l \vee c = l)$

**4.2 Four heuristics**
Effective identification of layout areas is based on four heuristics. The main steps of these heuristics are always the same:

1. Identify labels.
2. Assign cells to the identified labels.
3. Group them into equivalence classes.

The result of step one and step two are layout areas, step three makes a subsequent auditing possible, as the layout areas are grouped into equivalence classes [MC, 2002].

Every cell in the spreadsheet is assigned to either one label or to none. After the identification of one kind of label cells are immediately assigned. So in each step the number of assignable cells is reduced. The first two heuristics are based on the different kinds of labels, introduced in section 2.1. The third heuristic is based on the semantical assignment and the last heuristic is based on the geometrical assignment (see Definitions 6 and 7).

The four heuristics are processed in a specific order, namely in the order in which they are presented in the sequel. This accounts from the fact that each heuristic has a different significance for denoting semantic relatedness. So the first one has highest significance, the last one the lowest.

There is also an special order within the identification of labels. This is necessary if there are, for example, running numbers and counters existing in a way, that they enclose a block of cells. In such a case the question is, to what kind of label should the enclosed cells be assigned – to the running numbers or to the counters? Because of this problem a hierarchy is introduced between the five kinds of labels. This means, the identification of labels is done in the order in which the labels are introduced in section 2.1. After the identification of one kind of label, cells are immediately assigned and the co-occurrence of two different kinds of cells will not lead to non-determination.

The hierarchy, as proposed in this paper, was introduced after analysing some 40 spreadsheets. It reflects the significance of the different kinds of labels for denoting semantical relatedness. Most of the spreadsheets studied are freely available at [Decisioneering, 2004].

It is also possible that labels of the same kind describe a rectangular area, as they enclose a group of cells. In such a case it doesn't matter to which labels the cells are assigned, either to the vertical ones or the horizontal ones, as the labels are equivalent.

To provide auditing for the resulting layout areas, they are grouped into equivalence classes (see
[MC, 2002]). These equivalence classes can be checked for irregularities, which might be symptoms of errors. The layout areas have also been integrated in the existing prototype (see [Clermont, 2003]).

**Assignment Based on Running Numbers and Counters (Heuristic 1)**
The individual steps of this heuristic are

- the identification of running numbers, labels interpretable as ordinal numbers and counters (in this order),
- the geometrical assignment of cells to the identified labels. As mentioned above, this is done immediately after one of the three groups of labels is identified.

The result of these two steps are layout areas. Subsequently

- the layout areas are grouped into semantic classes and
- heuristic 2 is applied.

Figure 1 shows an example. In this example two different kinds of labels can be identified: running numbers (vertical running) in the cells A19:A22 and horizontal running counters in the cells B18:E18. In this example the two kinds of labels are arranged in a way that they enclose a block of cells. Now the importance of the stated hierarchy can be shown. If no hierarchy is made an assignment would not be possible, because the cells B19:E22 can be assigned to the running numbers as well as to the counters. However, the running numbers have higher priority.

As stated in our introduction, layout areas make an implicit identification of the orientation of the spreadsheet possible. The orientation of the example spreadsheet is row-wise as the running numbers are horizontal labels. In figure 1, four layout areas can be identified: A19:E19, A20:E20, A21:E21 and A22:E22. To highlight these layout areas, each layout area is marked with another greyscale.

To support auditing, these four layout areas are subsequently grouped into semantic classes. As this would go beyond the scope of this paper readers are referred to [Hipfl, 2004].

|    | A                   | B         | C         | D         | E         |
|----|---------------------|-----------|-----------|-----------|-----------|
| 16 | Production Forecast |           |           |           |           |
| 17 |                     |           |           |           |           |
| 18 | Year                | Quarter 1 | Quarter 2 | Quarter 3 | Quarter 4 |
| 19 | 2001                | 200.000   | 250.000   | 100.000   | 150.000   |
| 20 | 2002                | 150.000   | 220.000   | 120.000   | 145.000   |
| 21 | 2003                | 133.441   | 195.714   | 106.753   | 128.993   |
| 22 | 2004                | 118.711   | 174.109   | 94.969    | 114.754   |
| 23 | Total               | 602.152   | 839.823   | 421.722   | 538.747   |

Figure 1: Four layout areas build by means of running numbers.

**Assignment Based on Complete and Partial Identity (Heuristic 2)**
The individual steps of this heuristic are

- the identification of labels with complete identity and partial identity (in this order),
- the geometrical assignment of cells to the identified labels. As mentioned above, this is done immediately after one of the three groups of labels are identified.

The result of these two steps are layout areas. Subsequently

- the layout areas are grouped into semantic classes and
- heuristic 3 is applied.

Figure 2 shows an example. In this example complete identical labels can be identified. So each of the pairs of cells D7, H7 and E7, I7 and F7, J7 and G7, K7 contains complete identical labels. As these labels are horizontal labels, the orientiation of the example-sheet is column-wise.

Eight layout areas are identified: D7:D10, E7:E10, F7:F10, G7:G10, H7:H10, I7:I10, J7:J10 and K7:K10. In the example, each layout area built upon an identical label, has the same greyscale. These eight layout areas form four groups as areas with the same label belong together.

Figure 2: Eight layout areas build by means of complete identical labels

In a further step each of the four pairs of layout areas can be grouped into semantic classes. Further details are given in [Hipfl, 2004].

**Assignment Based on Semantics (Heuristic 3)**
The individual steps of this heuristic are:

For each label, not yet grouped to layout areas

- identify semantically assignable cells (see Definition 7) and
- geometrically assign the cells to the semantically related labels.

The result of these two steps are layout areas. Subsequently

- the cells within each layout area are checked for logical areas and
- heuristic 4 is applied.

Figure 3 shows an example. The formatting of the cells is described in the order *font*, *font size*, *font style*, and *font color*. The formatting of the block of cells B37:B40 is: *Arial, 10 pt., normal, black*. The formatting of the labels A37:A40 is *Arial, 12 pt., normal, black*. As the labels and the cells consist of different formatting a semantical assignment is not possible. Lets have a look at the other cells. The label A41 has the following formatting: *Arial, 12 pt., bold, black*. The cells B41:D41 have the same formatting. So an assignment is possible and the result is a layout area, highlighted in figure 3.

Figure 3: One layout area by means of semantical assignment

The processing of the result of this heuristic is different to those introduced before as the resulting layout areas are not grouped into semantic classes. In the first two heuristics the grouping to layout areas is based on the assumption, that similar labels are a sign for semantically related cells. But this heuristic is not based on the similarity of labels, so the resulting layout areas contain labels which might not be similar to each other. So the assumption made in heuristics 1 and 2, and consequently the grouping into semantic classes is not founded.

However, another abstraction technique can be applied. In [ACM, 2000] the method of logical areas is introduced. Cells within each layout area, resulting from this heuristic, can be checked for logical areas.

**Spatial Assignment (Heuristic 4)**

This last heuristic is in a way different from the others as the assignment is not based on any assumption of semantical relatedness. It

- identifies cells, which are not yet assigned and
- makes merely a geometrical assignment (see Definition 6).

The results of these two steps are layout areas that are not further processed.

The labels, as well as the cells, do not need to have any special characteristics. It also doesn't matter whether the labels to which the cells should be assigned, are already in another layout area. As this heuristic does not provide any statement on semantical relatedness, there is no benefit to group the result to semantic classes or logical areas.

This heuristic assigns cells which are not yet processed by a preceding step but described by a label to layout areas. After applying this heuristic only those cells are left that can not be unambiguously assigned to a label or which are not described by any label. Such cells can be an indication of errors or of bad design.

**5 CONCLUSION**

In this paper the spreadsheet visualization approach introduced by [Clermont, 2003] is extended by using layout information. The main assumption for the method of layout areas is that layout information is used by the spreadsheet-programmer to show the semantical relatedness of cells. On these grounds semantically related blocks of cells can be identified.

The aim of the extension is to relieve users from the need to be highly experienced in a specific spreadsheet visualization technique and it is reached as follows:

1. Explicit parametrization of the semantic classes' discovery algorithm is avoided.

    This is reached because semantic classes are identified basing on the layout areas. So the user does not need to specify spatial restrictions for blocks of cells with recurring semantics. In the extended approach labels and formatting information is

used to identify semantically related cells. Layout areas are built and subsequently out of them the semantic classes can be identified.

2. Implicit identification of the orientation of blocks is possible.

   In this approach the extent of the identified labels is used to indicate the orientation of cells. If labels are running horizontal, they describe columns – thus, the orientation is column-wise. If labels are running vertical, they describe rows – so, the orientation is row-wise. The orientation of the cells does not need to be explicitly stated.

3. The orientation is identified for each block of cells individually, not for the whole spreadsheet.

   Each layout area is build out of a set of labels which describe specific blocks of cells. These labels indicate the orientation of the blocks of cells they describe. Hence, the orientation is given only for blocks of cells.

4. It is possible to identify semantically related blocks of cells with different sizes.

   The size of a block of cells is not taken into consideration for the construction of layout areas. The cells are assigned to labels until the termination condition is reached. So, it is possible that two belonging layout areas consist of a different number of cells.